\documentclass[aps, prd, twocolumn, amsmath, floats,floatfix, superscriptaddress, nofootinbib]{revtex4}
\usepackage{graphicx}% Include figure files
\usepackage{bm}% bold math
\usepackage{hyperref}% add hypertext capabilities
\usepackage[utf8]{inputenc}
\makeatletter\def\Hy@Warning#1{}\makeatother
\hypersetup{
  colorlinks=true,        % false: boxed links; true: colored links
  linkcolor=blue,         % color of internal links
  citecolor=cyan,         % color of links to bibliography
  urlcolor=cyan            % color of external links
}
\usepackage{amsmath,amsfonts,amssymb}
\usepackage{times}
\usepackage[T1]{fontenc}
\usepackage{float}
\usepackage[normalem]{ulem}
\usepackage{multirow}
\usepackage{soul}

\newcommand{\pr}[1]{\ensuremath{\left[#1\right]}}
\newcommand{\pc}[1]{\ensuremath{\left(#1\right)}}
\newcommand{\chav}[1]{\ensuremath{\left\{#1\right\}}}

\DeclareMathOperator*{\argmin}{arg\,min} % thin space, limits underneath in displays
\newcommand{\be}{\begin{equation}}
\newcommand{\ee}{\end{equation}}
\newcommand{\bea}{\begin{eqnarray}}
\newcommand{\eea}{\end{eqnarray}}

\DeclareMathOperator{\E}{\mathbb{E}}
\makeatletter
\newcommand*{\rom}[1]{\expandafter\@slowromancap\romannumeral #1@}
\makeatother

\begin{document}
\title{From NS observations to nuclear matter properties: a machine learning approach}

\author{Valéria Carvalho}
\email{val.mar.dinis@uc.pt}
\affiliation{CFisUC, 
	Department of Physics, University of Coimbra, P-3004 - 516  Coimbra, Portugal}

\author{Márcio Ferreira}
\email{marcio.ferreira@uc.pt}
\affiliation{CFisUC, 
	Department of Physics, University of Coimbra, P-3004 - 516  Coimbra, Portugal}
	
\author{Constança Providência}
\email{cp@uc.pt}
\affiliation{CFisUC, 
	Department of Physics, University of Coimbra, P-3004 - 516  Coimbra, Portugal}

\date{\today}

\begin{abstract} 
This study is devoted to the inference problem of extracting the nuclear matter properties directly from a set of mass-radius observations. 
We employ Bayesian neural networks (BNNs), which is a probabilistic model capable of estimating the uncertainties associated with its predictions. To simulate different noise levels on the $M(R)$ observations, we  create three different sets of mock data. 
Our results show BNNs as an accurate and reliable tool for predicting the nuclear matter properties whenever the true values are not completely outside the training dataset statistics, i.e., if the model is not heavily dependent on its extrapolating capacities. Using real mass-radius pulsar data, the model predicted, for instance,  $L_{\text{sym}}=39.80\pm17.52 $ MeV and $K_{\text{sym}}=-101.67\pm62.86 $ MeV ($2\sigma$ interval).
Our study provides a valuable inference framework when new NS data becomes available.

\end{abstract}

\keywords{Bayesian neural networks, equation of state, nuclear matter, neutron stars}
\maketitle
\definecolor{vc}{rgb}{0.0, 0.5, 0.69}
\section{Introduction}
The properties and composition of matter properties inside neutron stars (NSs) remain an open question. The equation of state (EOS) of dense and asymmetric nuclear matter, realized inside NSs, is the key quantity of research in NS physics. Constraints on the EOS at moderate and high baryonic densities are mainly supplied by observations of massive NSs ($M/M_{\odot}$):  
$1.908 \pm 0.016 $ for PSR~J1614-2230   \cite{Demorest2010,Fonseca2016,Arzoumanian2017}, $2.01 \pm 0.04$ for
PSR~J0348 - 0432 \cite{Antoniadis2013}, $2.08 \pm0.07$ for PSR~J0740+6620 \cite{Fonseca:2021wxt}, and $2.13 \pm0.04$ for J1810+1714 \cite{Romani:2021xmb}. 
Furthermore, the recent advent of multi-messenger astrophysics combined different sources of information regarding NS physics (e.g., gravitational waves, photons, and neutrinos).
The observation of compact binary coalescence events such as GW170817 \cite{Abbott:2018wiz} and GW190425 \cite{Abbott:2020khf} by LIGO/Virgo collaboration posed further constrains on the EOS. The recent inferences of the
PSR J0030+045 pulsar mass \cite{Riley_2019,Miller19} and PSR J0740+6620 radius \cite{Riley2021,Miller2021,Raaijmakers2021}   by NICER (Neutron star Interior Composition ExploreR) experiment 
also narrowed down the possible NS matter scenarios. 
Observations from future experiments, with higher precision measurements,  such as the enhanced X-ray Timing and Polarimetry mission (eXTP) \cite{eXTP,eXTP:2018anb}, the (STROBE-X) \cite{STROBE-X}, and Square Kilometer Array \citep{SKA} telescope, will be fundamental in restricting the different scenarios for the NS matter properties.\\

The low density region of the EOS is constrained by chiral effective field theory (cEFT) \cite{Hebeler2013,Drischler:2017wtt}, and at high densities, pQCD is reliable (for a review \cite{ghiglieri2020perturbative}). The inference of the EOS of NSs given a set of theoretical and/or observational constraints is normally implemented via Bayesian inference scheme, see, e.g., \cite{Steiner:2010fz,Malik:2022zol,Malik:2023mnx}. Another interesting method also implemented is the Gaussian processes \cite{Essick2019,Landry:2020vaw,han2021bayesian,Gorda:2022jvk}. Let us point out that several approaches have considered agnostic descriptions of the EoS subjected to the above constraints, in particular,  in several works the two extreme constraining 
EoS are connected using a piecewise polytropic,  speed of sound or spectral interpolation,
and if necessary, causality is imposed \cite{Lindblom2012,Kurkela:2014vha,Most:2018hfd,LopeOter:2019pcq,Annala2019,Annala:2021gom}. These descriptions do not allow, however, for the determination of the nuclear matter properties.\\

The use of neural networks (NNs) based inference frameworks in high energy physics has gathered significant attention across various disciplines, as highlighted by the comprehensive review  \cite{Zhou:2023pti}. Our recent contribution \cite{Carvalho:2023ele} focused on employing a Bayesian neural network (BNN) to map NS observations to the speed of sound squared and proton fraction inside NSs. Additionally, we explored NNs in mapping the EOS of $\beta$-equilibrium NS matter to the properties of nuclear matter in \cite{ferreira2022extracting}, where we tested our final model with 31 nuclear models. Expanding beyond our contributions, several studies have been conducted \cite{ferreira2021unveiling,Fujimoto_2021,Fujimoto_2018,Fujimoto_2020,Soma:2022qnv,morawski2020neural,Krastev_2022,krastev2023deep,PhysRevC.106.065802,han2021bayesian,Thete:2023aej, Goncalves:2022smd}. For instance, deep NNs is explored to deduce nuclear matter properties in \cite{Krastev_2022,krastev2023deep} using two architectures: one mapping $M(R)$ curves to the EoS and the other mapping the EoS to nuclear matter properties. However, the need for robust uncertainty quantification is in need in both articles. Addressing questions like "How confident is a model about its predictions?" remains pivotal.\\

The present work aims to directly map a set of NS observables (mass-radius observations) to nuclear matter properties using an inference framework based on BNNs. Training on a dataset of relativistic mean field (RMF) models constrained by minimal conditions such as low-density properties and pure-neutron matter, our trained model facilitates instantaneous inference for diverse nuclear matter properties across various observation sets. This approach eliminates the need for separate Bayesian analyses for each observation set.\\

The paper is organized as follows. A basic introduction to BNNs is presented in Sec \ref{bnn}. Then the family of nuclear models chosen and the generation of the mock observation datasets is explained in Sec. \ref{dataset}. The model results for the properties of nuclear matter are discussed in Sec. \ref{results}, where the section is further divided into three tests, and lastly, the conclusions are drawn in Sec. \ref{conclusions}.

\section{Bayesian Neural Networks \label{bnn}}

Bayesian neural networks (BNNs) represent a Bayesian approach framework designed to quantify both aleatoric and epistemic uncertainties within a dataset \cite{jospin2022hands}. They can be viewed as an extension of traditional neural networks (NNs) wherein the model weights are stochastic and follow a specified probability distribution. The prediction of BNNs, denoted as $\bm{y^*}$ for a given input $\bm{x^*}$, transforms into a Bayesian model average

\begin{equation} \label{eq:baye_pred}
    P(\bm{y^*}|\bm{x^*}, D) = \int_{\bm{\theta}} P(\bm{y^*}|\bm{x^*},\bm{\theta})P(\bm{\theta}|D) d \bm{\theta},
\end{equation}
where $\bm{\theta}$ denotes the model's weights, $D$ the dataset, 
$P(\bm{y^*}|\bm{x^*},\bm{\theta})$ is the network's distribution (which captures the noise present in the data), and $P(\bm{\theta}|D)$ is the posterior distribution of our weights (which characterizes the model uncertainty). Both aleatoric and epistemic uncertainties are captured by, respectively, 
$P(\bm{y^*}|\bm{x^*},\bm{\theta})$ and $P(\bm{\theta}|D)$ (see \cite{olivier2021bayesian} for details). \\

Variational inference method is used to solve Eq.\ref{eq:baye_pred}, where the real posterior $P(\bm{\theta}|D)$
is approximated by a variational posterior $q_{\bm{\phi}}(\bm{\theta})$. The Kullback-Leibler relation, which measures the dissimilarity between two probability distributions, is used to quantify the quality of the approximation (i.e., how close is the variational posterior to the real posterior):
\begin{align}
   \text{KL}(q_\phi(\bm{\theta})||P(\bm{\theta}|D))
   &= \int_{\bm{\theta}} q_\phi(\bm{\theta}) \log \left( \frac{q_\phi(\bm{\theta})}{P(\bm{\theta}|D)}\right) d\bm{\theta}.
\end{align}
The variational posterior is determined by optimization the following objective function (see \cite{Carvalho:2023ele} for details):
\begin{align}
q_{\phi^*}&=\argmin_{q_{\phi}}\pr{\text{KL}(q_\phi(\bm{\theta})||P(\bm{\theta})) - \E_{q_\phi(\bm{\theta})}(\log P(D|\bm{\theta}))}.
\end{align}

\begin{figure*}[!hbt]
    \centering
\includegraphics[scale=0.4]{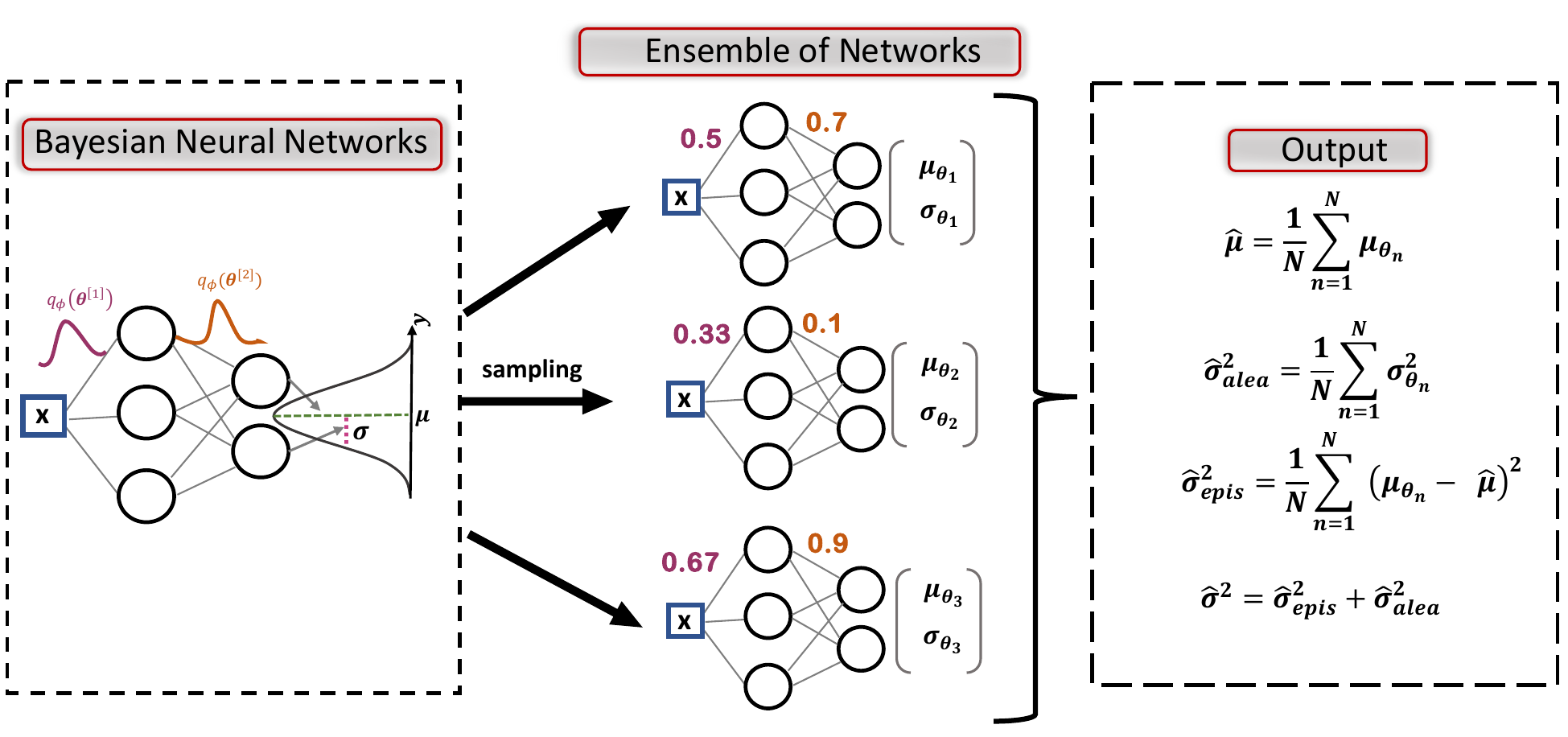}
    \caption{Intuitive illustration highlighting the predicted uncertainty quantification obtained by the optimized BNN model. The mean vector $\hat{\bm{\mu}}$ and the variance vector $\hat{\bm{\sigma}}^2$ of the predicting distribution $P(\bm{y^*}|\bm{x^*}, D)$ are calculated, for a fixed $\bm{x^*}$, utilizing the law of total expectation and total variance. These principles are elucidated more comprehensively in our earlier work \cite{Carvalho:2023ele}. 
The equation governing $\boldsymbol{\hat{\mu}}$ describes the mean vector obtained through the law of total expectation. Simultaneously, the variance vector, denoted by $\boldsymbol{\hat{\sigma}}^2$ and obtained via the law of total variance, is intricately partitioned into two distinctive components: an epistemic component ($\boldsymbol{\hat{\sigma}}^2_{epis}$) and an aleatoric component ($\boldsymbol{\hat{\sigma}}^2_{alea}$).    }
    \label{fig:BNN scheme}
\end{figure*}

Choosing a multivariate Gaussian for the variational posterior, $q_\phi(\bm{\theta})=\mathcal{N}(\bm{\mu}_q,\bm{\Sigma}_q)$, and a multivariate Gaussian with diagonal covariance matrix for the prior, $P(\bm{\theta})=\mathcal{N}(\mathbf{0},\mathbf{I})$, and using Monte Carlo sampling to calculate the expected values of our target objective function, we obtain

\begin{align}
    F(D,\phi)&=\frac{1}{2D_s}\left[\left( -\log 
    \det(\bm{\Sigma}_q) \right)- k + \operatorname{tr}\left(\bm{\Sigma}_q\right) + (\bm{\mu}_q)^T(\bm{\mu}_q)\right]\nonumber\\
    &- \frac{1}{B}\sum_{i=1}^B\frac{1}{N}\sum_{n=1}^N \log P(\bm{y_i}|\bm{x_i},\bm{\theta}^{(n)}),\label{eq:final_loss}
\end{align}
where $\bm{\theta}^{(n)}$ are samples from the variational posterior, $q_\phi(\bm{\theta})$, $B$ is the number of points of the mini-batch, $D_s$ is the number of points of the training dataset, $k$ is the dimension of the identity matrix of the prior and $N$ is the number of samples we take off the variational posterior (we used $N=10^4$). Once the network is trained and the mean $\bm{\mu}_q$ and covariance matrix $\bm{\Sigma}_q$ are found, which best approximate the true posterior, we solve Eq.~\ref{eq:baye_pred} via Monte Carlo sampling,

\begin{align}
P(\bm{y^*}|\bm{x^*}, D) = &\int_{\bm{\theta}} P(\bm{y^*}|\bm{x^*},\bm{\theta})q_\phi(\bm{\theta}) d \bm{\theta} \\
=& \frac{1}{N} \sum_{n=1}^{N} P(\bm{y^*}|\bm{x^*},\bm{\theta}^{(n)}), \quad \bm{\theta}^{(n)} \sim q_\phi(\bm{\theta}).
\end{align}\label{eq:p}
The meaning of the last equation is schematically illustrated in Fig.~\ref{fig:BNN scheme}. Each sampled weights $\bm{\theta}^{(n)}$ corresponds to a specific NN, and the BNNs output is determined by the ensemble prediction.

\section{Dataset \label{dataset}}

\subsection{Nuclear models}
The data we use for training the BNN models consists of nuclear models based on a relativistic mean field (RMF) description, where the nucleon interaction is mediated through the exchange of scalar-isoscalar mesons, vector-isoscalar mesons, and vector-isovector mesons. The model details are omitted here but can be seen in \cite{Malik:2023mnx}. The dataset consisting of 25 287 nuclear models was obtained by imposing minimal constraints on several nuclear saturation properties, reproducing 2$M_\odot$ NSs, and a consistent low-density pure neutron matter with N$^3$LO calculations in chiral effective field theory (see Table \ref{tab1}).

\begin{table}[!hbt]
 \caption{Constraints on the binding energy per nucleon $\epsilon_0$, incompressibility $K_0$, and symmetry energy $J_{\rm sym}$ at the nuclear saturation density $n_0$ (with 1$\sigma$ uncertainties). The pressure of pure neutron matter (PNM) is considered at densities of 0.08, 0.12, and 0.16 fm$^{-3}$, obtained from a $\chi$EFT calculation \cite{Hebeler2013}.}
  \label{tab1}
      \centering
 \setlength{\tabcolsep}{5.5pt}
      \renewcommand{\arraystretch}{1.1}
\begin{tabular}{cccc}
\hline 
\hline 
\multicolumn{4}{c}{Constraints}                                                        \\ 
\multicolumn{2}{c}{Quantity}                     & Value/Band  & Ref   \vspace{0.1cm}\\ \hline
\multirow{3}{*}{\shortstack{NMP \hspace{0.cm} }}  {[}fm$^{-3}${]}  & $n_0$ & $0.153\pm0.005$ & \cite{Typel1999}    \\
                                                & $\epsilon_0$ & $-16.1\pm0.2$ & \cite{Dutra:2014qga}   \\
                \hspace{1cm}   {[}MeV{]}        & $K_0$           & $230\pm40$   & \cite{Shlomo2006,Todd-Rutel2005}    \\
                              & $J_{\rm sym, 0}$           & $32.5\pm1.8$  & \cite{Essick:2021ezp}   \\
                              
                               &                 &                &                                                   \\
  \shortstack{PNM \\ {[}MeV fm$^{-3}${]}}                  & $P(\rho)$       & $2\times$ N$^{3}$LO    & \cite{Hebeler2013}   \\
  &$dP/d\rho$&$>0$&\\
%                               &                 &                &                           &                           \\
\shortstack{NS mass \\ {[}$M_\odot${]}}        & $M_{\rm max}$   & $>2.0$     &  \cite{Fonseca:2021wxt}      \\ 

\hline 

% \thickhline
\end{tabular}
\end{table}

To a good approximation, the EOS of nuclear matter can be written in terms of the binding energy per nucleon $\epsilon(n, \delta)$  and can be decomposed into a symmetric and asymmetric part \cite{Bombaci:1991zz,PhysRevC.80.045806}:
\begin{equation}\label{eq:asym-sym}
     \epsilon(n, \delta) \approx \epsilon_{\text{SNM}}(n) + E_{\text{sym}}(n)\delta^2 + \cdots \; ,
\end{equation}
where $\delta= \pc{n_n - n_p}/n$ is the isospin asymmetry, $n_p$ and $n_n$ are the proton and neutron density, respectively, $n=n_p+n_n$ is the baryonic density, $\epsilon_{\text{SNM}}(n) =\epsilon(n, \delta = 0) $ represents the binding energy per nucleon of symmetric nuclear matter, and $E_{\text{sym}}(n)$ denotes the symmetry energy, which could be written as
\begin{equation}\label{eq:Esym}
    E_{\text{sym}}(n)=\left. \frac{\partial ^2 \epsilon \pc{n,\delta}}{2 \partial \delta^2}\right|_{\delta=0}\; .
\end{equation}
Expanding both symmetric and asymmetric parts in a Taylor series around the saturation density $n_0$ until third order, we get 

\begin{align}
    \label{eq:E_total}
    \epsilon(n, \delta) &\approx \underbrace{ \epsilon_0 + \frac{K_0}{2}x^2+ \frac{Q_0}{6}x^3 }_{\epsilon_{\text{SNM}}(n)}+ \\ \nonumber 
    &\underbrace{  \pc{J_{\text{sym}} +   L_{\text{sym}}x + \frac{K_{\text{sym}}}{2}x^2 + \frac{Q_{\text{sym}}}{6}x^3}}_{E_{\text{sym}}(n)}\delta^2 + \cdots \; , \nonumber
\end{align}

where $x=(n-n_0)/(3n_0)$.
The $\epsilon_0=\epsilon_{\text{SNM}}(n_0)$ term denotes the binding energy per particle, and $J_{\text{sym}}$ the symmetry energy, $K_0$ is the incompressibility coefficient, $Q_0$ the skewness, while  $L_{\text{sym}}$, $K_{\text{sym}}$, $Q_{\text{sym}}$ are, respectively, the slope, curvature and skewness of symmetry energy at saturation density respectively. 
The dataset statistics for the nuclear matter properties are in Table \ref{tab:describe}. 
A detailed examination of these parameters, including asymmetric parameters with higher order effects in Eq. \ref{eq:asym-sym}, can be found in \cite{Tovar2021}.

\begin{table}[!hbt]
\caption{Extreme values and respective mean and standard deviation for nuclear matter parameters from the source dataset we are employing. The dataset contains a total of 25287 nuclear
models. }
\begin{tabular}{cccccc}
\toprule
\multicolumn{2}{c}{Y}  &          $\overline{Y}$ &          $\sigma_Y$ &           $Y_{min}$ &         $Y_{max}$ \\ \hline
$n_0$  & [fm$^{-3}$] &    0.153351 &     0.003356 &      0.140047 &       0.166157 \\
$E_0$    & &   -16.099182 &     0.176263 &    -16.862894 &     -15.251849 \\
$K_0$   &   & 269.555070 &    19.696655 &    193.038134 &     347.284418 \\
$Q_0$    &   &-381.870901 &    60.604233 &   -617.004518 &    -15.088353 \\
$J_{\text{sym}}$ & [MeV]  &  31.870120 &     1.466915 &     25.416222 &       37.963163 \\
$L_{\text{sym}}$ &    & 41.412327 &     9.824848 &     21.789640 &     93.972049 \\
$K_{\text{sym}}$ &  & -107.711394 &    31.334004 &   -212.745127 &  16.062865 \\
$Q_{\text{sym}}$ &  & 1274.851388 &   297.082380 &    -63.828371 &    1962.652643 \\\hline
\end{tabular}
\label{tab:describe}
\end{table}

\subsection{Structure}
Our objective involves utilizing simulated observations of neutron stars (referred to as the set $\bm{X}$) as inputs for a BNN model, denoted as $P(\bm{Y} | \bm{X}, \bm{\theta})$, where $\bm{\theta}$ represents parameters sampled from the variational posterior distribution of the model, i.e., $\bm{\theta} \sim q_\phi(\bm{\theta})$. The primary objective is to establish a probability distribution for the output space, denoted as the set $\bm{Y}$. This set encompasses properties of nuclear matter, introduced in the preceding section and detailed in Table \ref{tab:describe}.
However, our focus is specifically on training and predicting the incompressibility $K_0$ and skewness $Q_0$ of symmetric nuclear matter, and the symmetry energy $J_{\text{sym}}$, along with its slope $L_{\text{sym}}$, curvature $K_{\text{sym}}$, and skewness $Q_{\text{sym}}$. 
We are not predicting the values of $n_0$ and $E_0$ because they are theoretical and experimental well restricted and have been strongly constrained when the EoS dataset was generated \cite{Malik:2023mnx}. Furthermore, those two variables show small correlations with the input of our model, being less informative for the predictions. %\mf{[What more can we say/justify on this choice? CP]}
In a similar work in \cite{krastev2023deep}, the author also kept constant the values of $E_0$ and $n_0$, doing a study with the use of two NN architectures which first maps the observations to the EOS curve and then maps pressure times baryonic density to the parameters of  nuclear matter. \\\\

The input $\bm{X}$ and output $\bm{Y}$ set are composed of $\mathcal{D}$ rows of vectors $\bm{x}$ and $\bm{y}$, respectively, where $\mathcal{D}$ represents the dataset size we are utilizing. In simpler terms, $\bm{Y}$ is expressed as $\bm{Y}= \{\bm{y}^{(i)}\}_{i=1}^\mathcal{D}$, and $\bm{X}$ is denoted as $\bm{X}= \{\bm{x}^{(i)}\}_{i=1}^\mathcal{D}$. The size of the output vector is $ \bm{y}^{(i)}= \{K_0,Q_0, J_{\text{sym}}, L_{\text{sym}}, K_{\text{sym}},Q_{\text{sym}}\}$, while the input vector $\bm{x}^{(i)}$ is defined by 10 values: $\bm{x}^{(i)}=\{(M_j^{(i)},R_j^{(i)})\}_{j=1}^{5} $, representing five simulated observations $M_j(R_j)$ in the mass-radius curve.
Concerning the dataset size, we randomly divided the total EOS models into two distinct sets: a training set denoted as $\bm{X}$ and $\bm{Y}$, incorporating 80\% of the data ($\mathcal{D}=$22758 EOS), and a test set denoted as $\bm{X}_T$ and $\bm{Y}_T$, comprising the remaining 20\% ($\mathcal{D}_T=$2529 EOS). In summary, our output elements are 6-dimensional vectors denoted as $\bm{y}^{(i)}$ with a total of training and test data expressed as $\bm{Y}=\{\bm{y}^{(i)}\}_{i=1}^{22758}$ and $\bm{Y}_T=\{\bm{y}^{(i)}\}_{i=1}^{2 529}$, respectively. Regarding the input space, we have a 10-dimensional vector $\bm{x}^{(i)}= \{(M_j^{(i)},R_j^{(i)})\}_{j=1}^{5} $, and both the training and test data are denoted as $\bm{X}=\{\bm{x}^{(i)}\}_{i=1}^{22758}$ and $\bm{X}_T=\{\bm{x}^{(i)}\}_{i=1}^{2 529}$.\\\\

\subsection{Observation mock data\label{generation}}

\begin{figure*}[!hbt]
    \centering
    \includegraphics[width=0.9\linewidth]{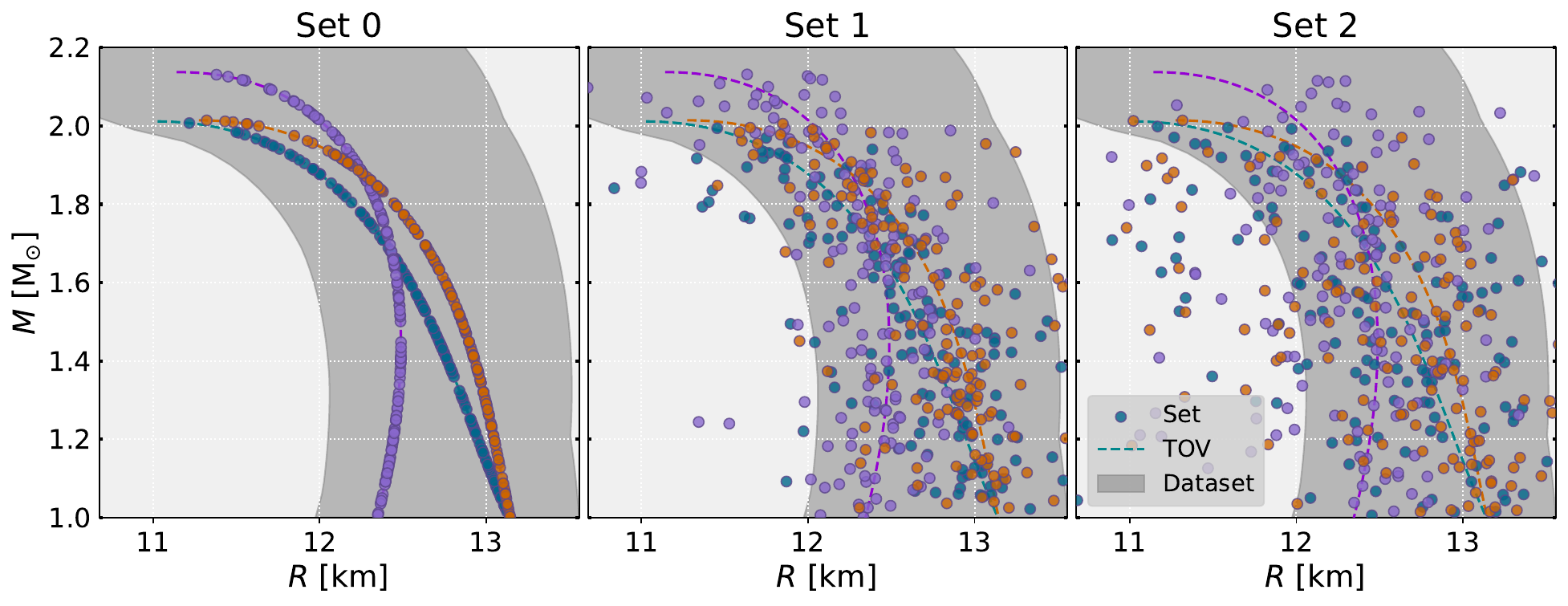}
    \caption{The grey area represents the extremes of the dataset, while the three lines represent three different EOSs. The points along the lines represent the values corresponding to the respective sets, with the dataset size reduced to $n_s=30$ for a clearer visualization.}
    \label{fig:3_sets}
\end{figure*}

The statistical procedure for generating the five mock observations $M_j(R_j)$ in the mass-radius curve with distinct input noises originating in different datasets unfolds through the following steps. For each EOS,
\begin{enumerate}
    \item We randomly sample 5 NS mass values, $M_j^{(0)}$, from a uniform distribution within the range of $1.0M_{\odot}$ to $M_{\text{max}}$. 
    \item We sample 5 values, $\sigma_{j,M}$, from a uniform distribution spanning the interval $[0, \sigma_M]$ ($\sigma_M$ is in Table \ref{tab:sets}).
    \item We sample 5 values, $\sigma_{j,R}$, from a uniform distribution spanning the interval $[0, \sigma_R]$ ($\sigma_R$ is in Table \ref{tab:sets}).
    \item  The NS radius, $R_j$, is sampled from a Gaussian distribution centered at the TOV solution, denoted as $R(M_j^{(0)})$, with the standard deviations $\sigma_{j,R}$.
    \item Lastly, the NS mass is sampled from a Gaussian distribution centered at $M_j^{(0)}$ with the standard deviations  $\sigma_{j,M}$.
\end{enumerate}
\noindent This process is summarized by the following equations:

  \begin{align}
   M_j^{(0)} &\sim \mathcal{U}{[1,M_{\text{max}}]}\quad (\text{in units of  }\textup{M}_\odot)\\
        R_j &\sim \mathcal{N}\left(R\left(M_j^{(0)}\right),\sigma_{j,R}^2 \right),  \text{ where }  \sigma_{j,R} \sim \mathcal{U}{[0,\sigma_R]}\\ 
    M_j &\sim \mathcal{N}\left(M_j^{(0)},\sigma_{j,M}^2 \right),  \text{ where }     \sigma_{j,M} \sim \mathcal{U}{[0,\sigma_M]} \quad j=1,...,5.  
 \end{align}
This method is very similar to the method implemented in \cite{fujimoto2021extensive}.
By performing the above numerical procedure, we construct the initial input structure $\bm{x}=[M_1,\cdots,M_5,R_1,\cdots,R_5]$, where each pair ($M_j,R_j$) is a single realization of the above procedure ({\it observation}). These pairs collectively characterize the $M(R)$ diagram of a given EOS. We will replicate $n_{\text{s}}$ times the aforementioned procedure for the same EOS, i.e., $n_{\text{s}}$ is the number of times we resample the input vector, $\bm{x}$, for the same EOS. As a result, the dataset is organized as $\mathbb{X}=\{\bm{X}_i\}_{i=1}^{n_s}$ and $\mathbb{Y}=\{\bm{Y}_i\}_{i=1}^{n_s}$. This approach expands our dataset to a size of $\mathbb{D}=n_s \times \mathcal{D}$, where $\mathcal{D}$ denotes the original dataset size for the number of EOSs. For example, by choosing  $n_{\text{s}}=120$ implies the repetition of these procedures 120 times for each EOS, resulting in $\mathbb{X}=\{\bm{X}_i\}_{i=1}^{120}=\chav{\bm{X_1},\bm{X_2},\cdots,\bm{X_{120}}}$.\\
We have generated a total of three distinct datasets, whose properties are detailed in Table \ref{tab:sets}. The standard deviation for set 2 was chosen by considering the 10 initial values for the standard deviations of mass and radius in Table \ref{tab:observat}, which correspond to the ones with the smallest standard deviation for the radius. We calculated the mean of these 10 values. Set 1 comprises half the values of set 2, and set 0 represents the values without noise, as illustrated in Fig.~\ref{fig:3_sets}. 
The reason we only chose the 10 values with a lower standard deviation of the radius is that otherwise, the input noise would be so high that the model would not be able to extract anything from it.
The creation of these three sets aims to investigate how the model responds to different levels of noise. We utilized 120 mock observations ($n_s=120$) in the training sets for each EOS, while $n_s=1$ was employed for the test sets. This distinction mimics a real-world scenario where access to a single mock observation of the "true" EOS is typical. Here, a single mock observation corresponds to $n_s=1$, representing five $M_j(R_j)$ mock observations. It's essential to note that, for each EOS, there are 600 points in the $M(R)$ diagram, comprising 120 simulated observations for 5 neutron star mock observations each. 

In our previous investigation \cite{Carvalho:2023ele}, we employed $n_{\text{s}}=60$ during the training phase. However, for the current study, where we have reduced the dimension of the output vector, we wanted to enhance the model's performance without compromising training time and RAM usage. After experimenting with different adjustments to either the input vector or the $n_{\text{s}}$ value for the training, we found that setting $n_{\text{s}}=120$ yielded the optimal balance between improved performance and computational efficiency.

\begin{table}[!hbt]
   \caption{Generation parameters for each dataset.  }
    \label{tab:sets}
    \begin{tabular}{ccc}
    \toprule
    Dataset & $\sigma_M \;[ M_\odot]$ & $\sigma_R \;$ [km]   \\ \hline
    0  & 0     & 0   \\ \hline
    1  &  0.136   & 0.626     \\ \hline
    2  &   0.271   & 1.253     \\ \hline
    \end{tabular}
\end{table}

\subsection{Training procedure} \label{subsection}

To investigate the response of BNNs to varying input noises and output targets, we conducted a series of experiments involving the training of diverse functional and stochastic models. The BNN models were trained using distinct datasets generated following the guidelines provided in Section \ref{generation}. In the training stage, a portion of the training data was randomly chosen as a validation set, with 80\% dedicated to actual training and 20\% for validation. Additionally, both the input data $\bm{X}$ and output data $\bm{Y}$ were standardized\footnote{$z=(x-\mu)/\sigma$, where $\mu$ stands for the mean of the training dataset and $\sigma$ is the standard deviation of the training dataset.}, given the significant differences in the ranges of the output vector values. Defining the functional models entailed adjusting the number of neurons, layers, and activation functions. Table \ref{tab:final_models2} shows the best functional model for all the datasets mentioned in Table \ref{tab:sets}.

In our exploration of hidden layers, we considered hyperbolic tangent, softplus, and sigmoid activation functions, with a linear activation function chosen for the output layer. The architecture incorporates three hidden layers, each consisting of ten neurons.
The output layer consistently contains 12 neurons, with 6 dedicated to representing the mean and another 6 for the standard deviation of the output probability distribution function. It is noteworthy that we intentionally avoided incorporating correlation in the output layer, as its inclusion led to inferior performance. Consequently, the output layer is exclusively focused on capturing the mean and standard deviation information of the output distribution. The architecture adopted in this study features two to three hidden layers, maintaining a consistent number of neurons within each hidden layer.

The most favorable outcomes are achieved by employing the sigmoid activation function in the hidden layers, ensuring minimal loss and preventing divergence. It was determined that the optimal number of hidden layers is three. Detailed information on these configurations is available in Table \ref{tab:final_models2}. During training, we use a learning rate of 0.001 and employ the ADAM optimizer \cite{kingma2014adam} with the AMSgrad improvement \cite{reddi2019convergence}. The models undergo training for 4000 epochs, utilizing a mini-batch size of 1536. \\

\begin{table}[!hbt]
\caption{Structures of the final BNN models.}
\centering
\setlength{\tabcolsep}{10pt}
\begin{tabular}{ccc}
\toprule
\textbf{Layers} & \textbf{Activation}& \textbf{Neurons}\\ \hline
Input & N/A & 10 \\ \hline
Hidden Layer 1 & Sigmoid & 10  \\ \hline
Hidden Layer 2 & Sigmoid & 10 \\ \hline
Hidden Layer 3 & Sigmoid & 10 \\ \hline
Output & Linear & 12 \\ \hline
\end{tabular}
\label{tab:final_models2}
\end{table}
Concerning the stochastic model, we employ a Gaussian prior with a mean of zero and a standard deviation of one. While this prior choice lacks a specific theoretical justification, it serves as a reasonable default prior, as discussed in \cite{jospin2022hands}. Future research could delve further into investigating the impact of prior parameters, similar to the approach taken in reference \cite{bollweg2020deep}. Additionally, we select a multivariate normal distribution as the variational posterior, initialized with a mean of 0 and a diagonal covariance matrix, where the standard deviation is equal to $\log(1+\exp{0}) = 0.693$.
Furthermore, all BNNs models were coded using Tensorflow library \cite{tensorflow2015-whitepaper}, more specifically we use Keras \cite{chollet2015keras}, an high-level API of the TensorFlow.

\section{Results \label{results}}

After training our model, we proceed to assess its performance through a series of tests. Initially, we evaluate the model's accuracy using the test set derived from the original set of EOS outlined in Section \ref{test_dataset}. Subsequently, in Section \ref{NM_aplication}, we examine the model's performance with different nuclear models. Lastly, we conduct a test on our model using the observations outlined in Section \ref{obs_aplica}. We should clarify that whenever we say, for instance, the results of set 1, we are specifically indicating the BNN model trained on dataset 1 and evaluated with a corresponding noise level from set 1. This applies to Sections \ref{test_dataset} and \ref{NM_aplication}. However, for the last section \ref{obs_aplica}, it's crucial to highlight that no noise is applied to the test set, i.e., the test set consists of the mean values of the pulsar observations.

\subsection{Test set prediction \label{test_dataset}}

Analyzing the predictions of our BNN model across the three sets, we begin by examining the behavior of model predictions over the entire test set. We define the normalized residuals as $\Gamma(Y)={(\hat{\mu}_Y-Y_T)}/\hat{\sigma}_Y$, where $\hat{\mu}_Y$ and $\hat{\sigma}_Y$ are the mean and standard deviation predicted by the model, and $Y_T$ is the True value. Subsequently, we evaluate the coverage probability.

\begin{figure}[!hbt]
    \centering
\includegraphics[width=0.9\linewidth]{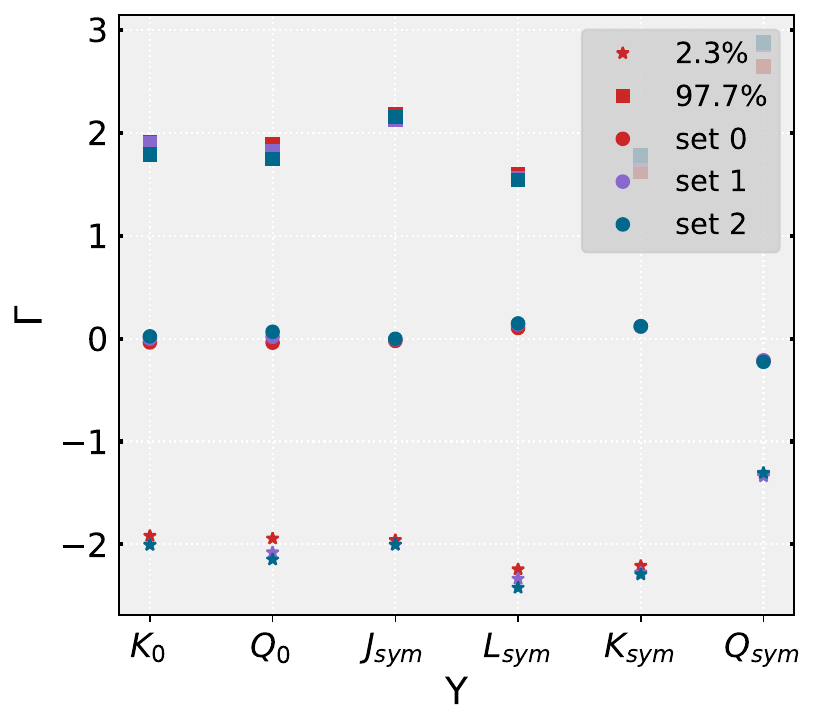}\\
\includegraphics[width=0.9\linewidth]{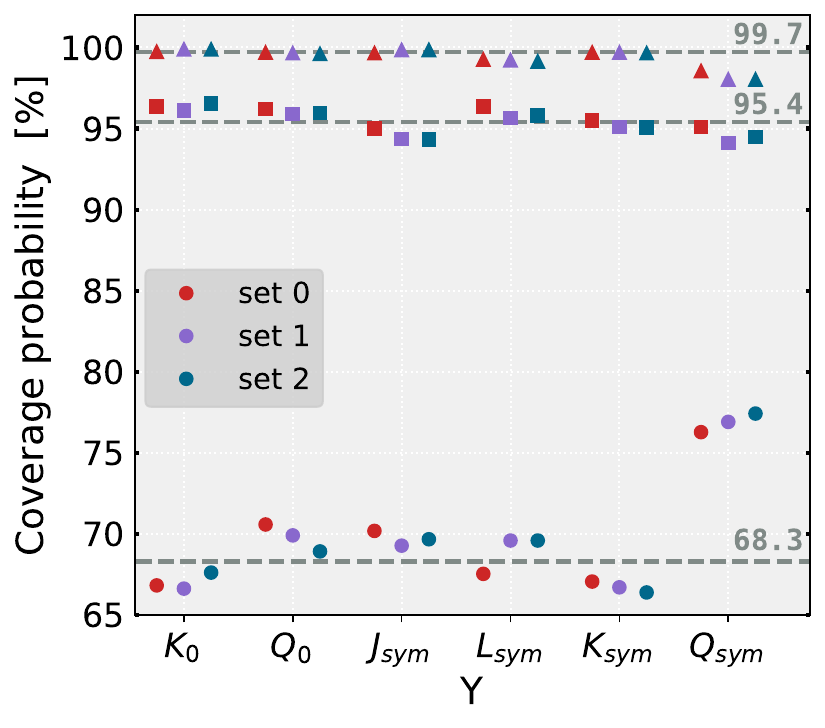}
    \caption{In the upper plot, the median for $\Gamma(Y) = {(\hat{\mu}_Y - Y_T)}/\hat{\sigma}_Y$ is denoted by circular markers ($\bullet$) for each set, while the 95\% confidence interval is indicated between square ($\blacksquare$) markers (97.7\% CI) and star ($\star$) markers (2.3\% CI). In the bottom plot, the coverage probability is represented with circular markers ($\bullet$) for 68.3\%, square markers ($\blacksquare$) for 95.4\%, and triangular markers ($\blacktriangle$) for 99.7\% confidence intervals.}
    \label{fig:gamma}
\end{figure}

The results for these two metrics are illustrated in Fig. \ref{fig:gamma} for our 6-dimensional output vector across the three sets.
Focusing on $\Gamma$ (upper panel), we observe that for all three datasets, approximately 50\% of the values cluster around zero. Additionally, the 95\% confidence interval is situated between $\pm 2\sigma$, with slight deviation for $Q_{\text{sym}}$.
The analyzis of the coverage probability (bottom panel) shows the model's ability to capture the data distribution. This metric quantifies the percentage of values within specific intervals relative to the total values in the test set. In particular, we examine whether the proportion of values falling within $1\sigma$ of the output distribution aligns with the expected 68.3\%. This evaluation is then extended to $2\sigma$ (95.4\%) and $3\sigma$ (99.7\%) intervals.
The model demonstrates an effective prediction of the associated percentage for the standard deviation being measured.

The observed behavior of $Q_{\text{sym}}$ in the coverage probability can be attributed to the considerable skewness in the distribution of $\Gamma$, distinguishing it from the other nuclear properties.

It is crucial to note from these two plots that the results for the three sets are remarkably consistent. This observation indicates that irrespective of the increased noise in our input vector, which leads to an increased uncertainty as evident in subsequent plots, the model consistently adapts to preserve the statistics of the dataset.\\

To gain a deeper insight into the impact of each set on the model prediction, we introduce the following metric:

\begin{equation}
  \eta\pr{a,b}=\frac{1}{T}\pc{\sum_{i=1}^{T}\frac{\hat{\sigma}_i^{a} - \hat{\sigma}_i^b }{\hat{\sigma}_i^a}} \times 100.
  \label{eq:eta}
\end{equation}

\begin{figure}[!hbt]
    \centering
    \includegraphics[width=0.9\linewidth]{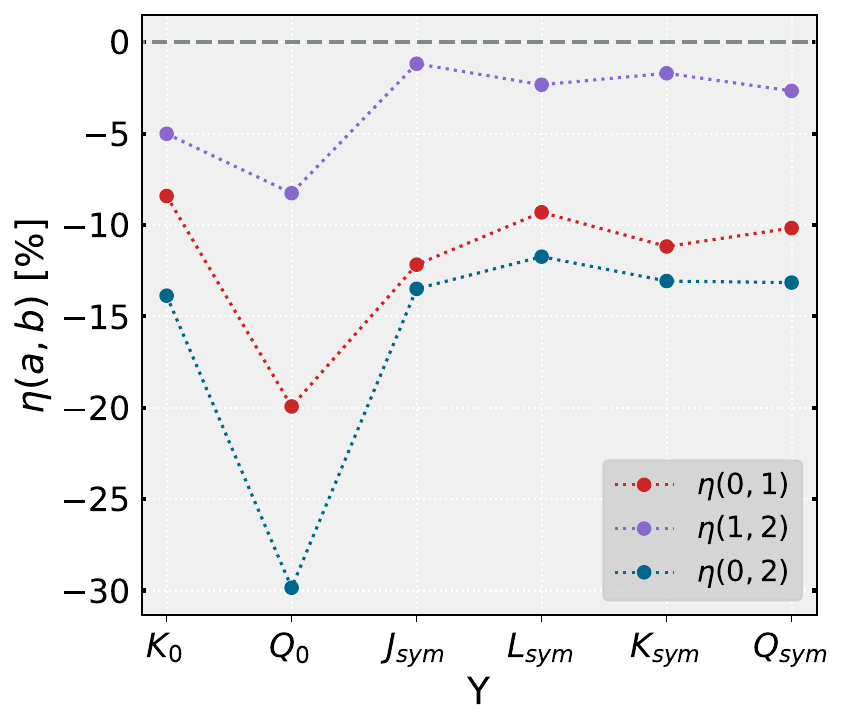}
    \caption{ Prediction uncertainty deviation $\eta(a,b)$ for the output vector for models a and b as defined in Eq.~\ref{eq:eta}. }
    \label{fig:eeta}
\end{figure}

This metric provides insights into the percentage uncertainty deviation between models trained with sets 'a' and 'b' across the 6 values of the output vector, and T is the total number of EOS in the test set. The results are shown in Fig.~\ref{fig:eeta}. Upon examining the plot, a notable observation is that, for all three $\eta$ vectors, the quantity $Q_0$ demonstrates the most significant variation.

Qualitatively this aligns with our expectations, given that $Q_0$ exhibits the highest correlation with the observables, as illustrated in Fig.~\ref{fig:corre} in Appendix \ref{ap:correl}. It's crucial to acknowledge that the correlation observed is only based on linear dependence. An alternative correlation measure, known as Kendall rank, is shown in Fig. 13 of the work \cite{Malik:2023mnx}, where once again, $Q_0$ stands out with the highest score for $R_{1.4}$ and $R_{\text{max}}$.  Additionally, we acknowledge the inherent challenge posed by the stochastic nature of our model, which contributes to the complexity of quantitatively interpreting the results. 

Moreover, the values of $\eta(0,1)$, $\eta(1,2)$, and $\eta(0,2)$ consistently exhibit negative values, as anticipated. This is attributed to the absence of input noise in set 0 and the larger noise in the input of set 2 compared to set 1. Examining the order of $\eta$ values, the blue curve exhibits a more negative trend, as expected, indicating a greater difference in input noise. 

Additionally, $\eta(0,1)$ appears more negative than $\eta(1,2)$, emphasizing the impact of varying the input noise from a uniform distribution for sets 1 and 2. Occasionally, these two sets converge to similar standard deviation values for the noise, a scenario less likely for set 0, which consistently lacks input noise. 

In the final examination of this section, we analyze the model's residuals, $\delta = {(\hat{\mu}_Y-Y_T)}$, by examining their mean and standard deviation, as detailed in Table \ref{tab:residuals}. Notably, the model exhibits higher accuracy for properties of nuclear matter with lower order, which is correlated with their smaller range of values. When comparing between sets, it becomes apparent that the standard deviation of residuals is greater for set 2, followed by set 1, and, finally, set 0. This observation indicates that the model achieves higher accuracy when trained with less input noise, aligning with our expectations. 

\begin{table*}[!hbt]
    \caption{The mean and the standard deviation of the model residuals $\delta = {(\hat{\mu}_Y-Y_T)}$ for the three sets. }
    \label{tab:residuals}
    \centering
    \begin{tabular}{cccc}\toprule
\multirow{2}{*}{Y \ [MeV]}  & &     \multicolumn{2}{c}{Set 0}     \\ \cline{3-4} 
&& $\overline{\delta}$& $\sigma_\delta$ \\
\hline
$K_0$ & &-0.663391 &   16.440592 \\
$Q_0$ && -1.879105 &   39.562439 \\
$J_{\text{sym}}$ && -0.002277 &    1.272795 \\
$L_{\text{sym}}$ &&  0.004316 &    8.769408  \\
$K_{\text{sym}}$ &&  0.157757 &   27.521557  \\
$Q_{\text{sym}}$ & & 2.541514 &  280.488010 \\ \hline
    \end{tabular}
    \begin{tabular}{ccc}\toprule
&      \multicolumn{2}{c}{Set 1}      \\ \cline{2-3} 
&$\overline{\delta}$& $\sigma_\delta$\\
\hline
&-0.505939 &  17.894562 \\
&-1.338307 &  48.660290 \\
 &0.003521 &   1.436066 \\
 &0.039834 &   9.523284 \\
&-0.311028 &  30.918154 \\
& 2.908357 & 291.857966 \\ \hline
    \end{tabular}
   \begin{tabular}{ccc}\toprule
&      \multicolumn{2}{c}{Set 2}      \\ \cline{2-3} 
&$\overline{\delta}$& $\sigma_\delta$\\ \hline
&-0.315610 &  18.705447 \\
&-0.381916 &  52.664798 \\
& 0.003998 &   1.454013 \\
& 0.021291 &   9.692659 \\
&-0.179070 &  31.494850 \\
 &2.560507 & 293.681477 \\ \hline
    \end{tabular}
\end{table*}

For instance, the $L_{\text{sym}}$ residuals standard deviation (MeV) for the model predictions is 8.77 (set 0), 9.52 (set 1), 9.69 (set 2). These uncertainties are not larger, and generally smaller than the ones obtained in several works where the uncertainties linked to experimental, observational or theoretical uncertainties are of the order 10-28\% \cite{Tsang:2012se,Lattimer2013,Oertel:2016bki}. 

\subsection{Nuclear models application \label{NM_aplication}}

\begin{figure}[!hbt]
    \centering
    \includegraphics[width=1\linewidth]{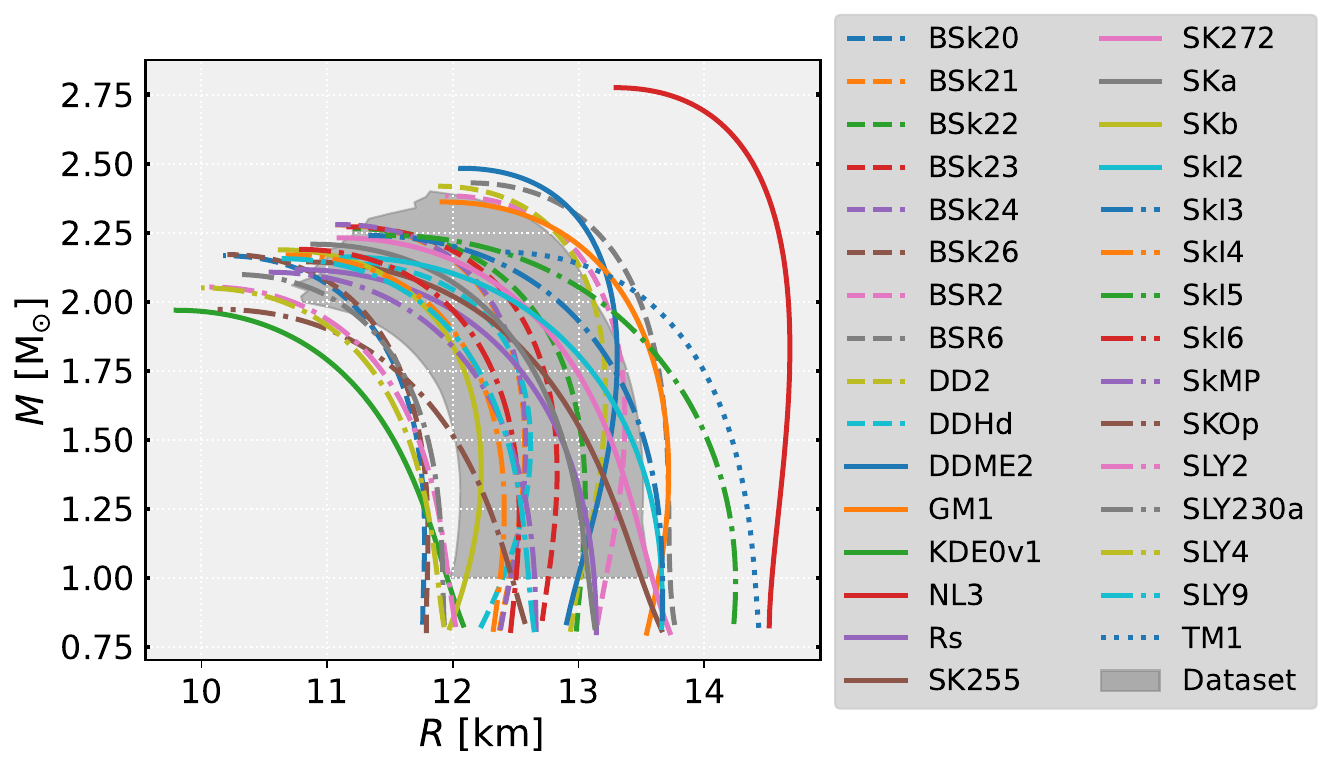}
    \caption{The grey band represents the dataset region without noise, and the multiple lines represent the 31 nuclear models.
    We have a total of 14 EoS inside the dataset region without noise for the input space and  12 EOS inside the dataset region for the output space.}
    \label{fig:M-R-nuclear}
\end{figure}

After confirming the ability of our model to predict the test set, we extend our assessment to evaluate its performance with additional nuclear models. Specifically, we consider a set of 31 unified EOS built from a relativistic mean field (RMF) approach and non-relativistic Skyrme interactions \cite{Fortin:2016hny}. These EoS have been considered because they are unified, meaning that the inner crust and the core have been obtained from the same model, and they span a large set of nuclear matter properties. This last characteristic will allow us also to test the extrapolation ability of our BNN model. Besides, this set has already been used in a previous study using ML methods \cite{ferreira2022extracting}, and, therefore some comparisons are possible.
The mass-radius curves for these models are represented in Fig. \ref{fig:M-R-nuclear}.

In Section~\ref{test_dataset} we have measured the performance of three BNN models, each one trained in a distinct dataset (see Table~\ref{tab:sets}). Here, we are applying these models to an independent set of nuclear models. For that, we first need to generate the corresponding observation sets (input data) with the same statistical properties as the ones in which the models were trained. For instance, if we want to apply the BNN model trained on set 0, we need to generate an observation set for these nuclear models without any noise (see Table~\ref{tab:sets}).
Furthermore, the generation of the observation sets follows a similar procedure used in Sec~\ref{generation}, albeit with some modifications. For each nuclear model and level noise, the procedure is as follows: i) randomly selecting 20 points from each $M(R)$ curve, ii) from these 20 points, 100 sets of 5 points are generated by sampling without replacement, resulting in 100 distinct input vectors $\boldsymbol{x}$ with 10 values each.

The selection of 20 points for the mass-radius $M(R)$ curve was driven by the number of available observations, as indicated in Tab. \ref{tab:observat}, which is 18. Given the relatively small size of the input vector, consisting of only 5 pairs, we opted to generate the 100 subsets for each nuclear model.

\begin{figure*}[!hbt]
    \centering
    \includegraphics[width=0.9\linewidth]{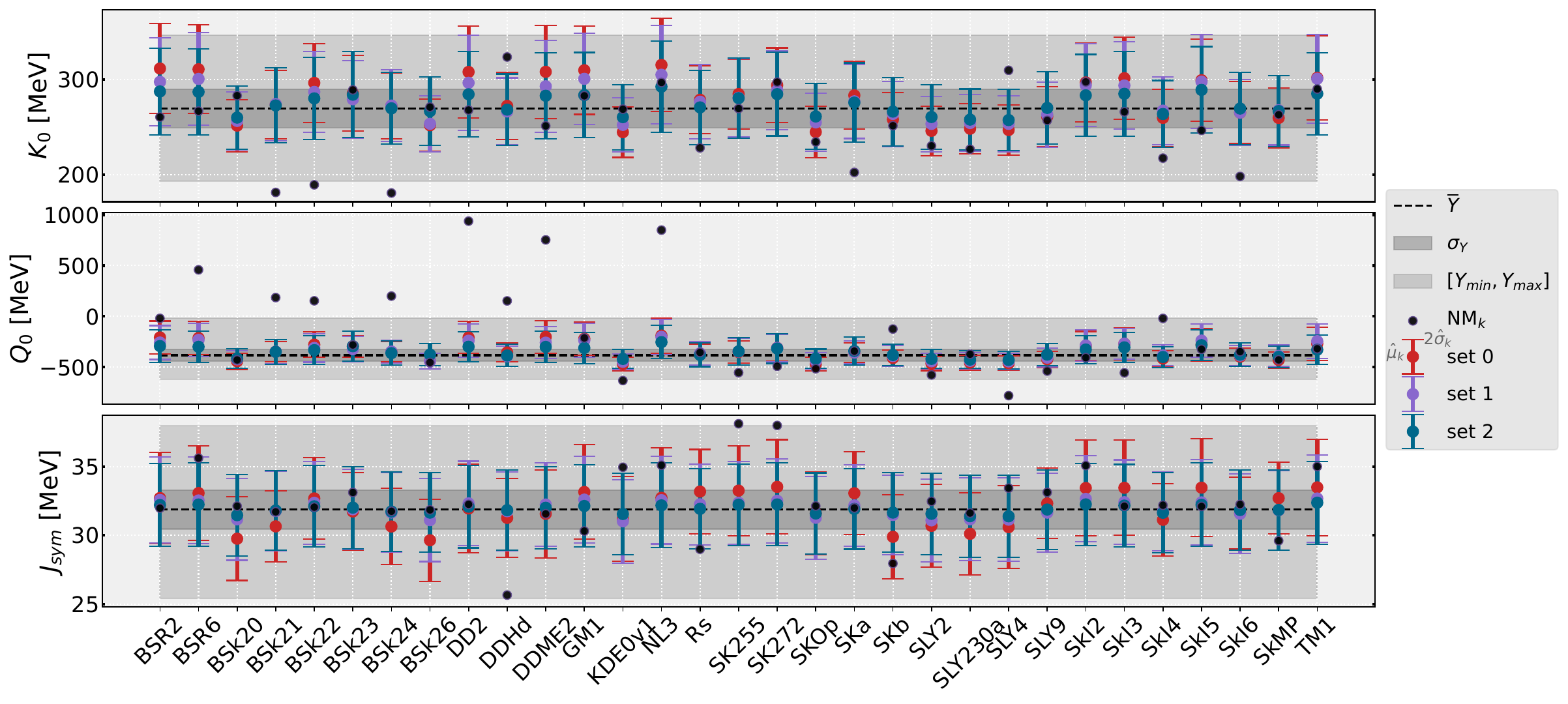}\\
    \caption{
    Representation of the $K_0,\, Q_0,\, J_{\text{sym}}$ output values, where the light grey band illustrates the minimum and maximum values within the training dataset [$Y_{\text{min}},Y_{\text{max}}$]. The dashed line and dark grey band correspond, respectively, to the mean and the $1\sigma$ CI of the training region ($\overline{Y}\pm\sigma_Y$). The error bar denotes the $2 \sigma$ range prediction by the BNN model ($\hat{\mu}_k \pm 2\hat{\sigma}_k$) for the sets 0, 1, and 2.}
    \label{fig:NM_1}
\end{figure*}

\begin{figure*}[hbt!]
    \centering
    \includegraphics[width=0.9\linewidth]{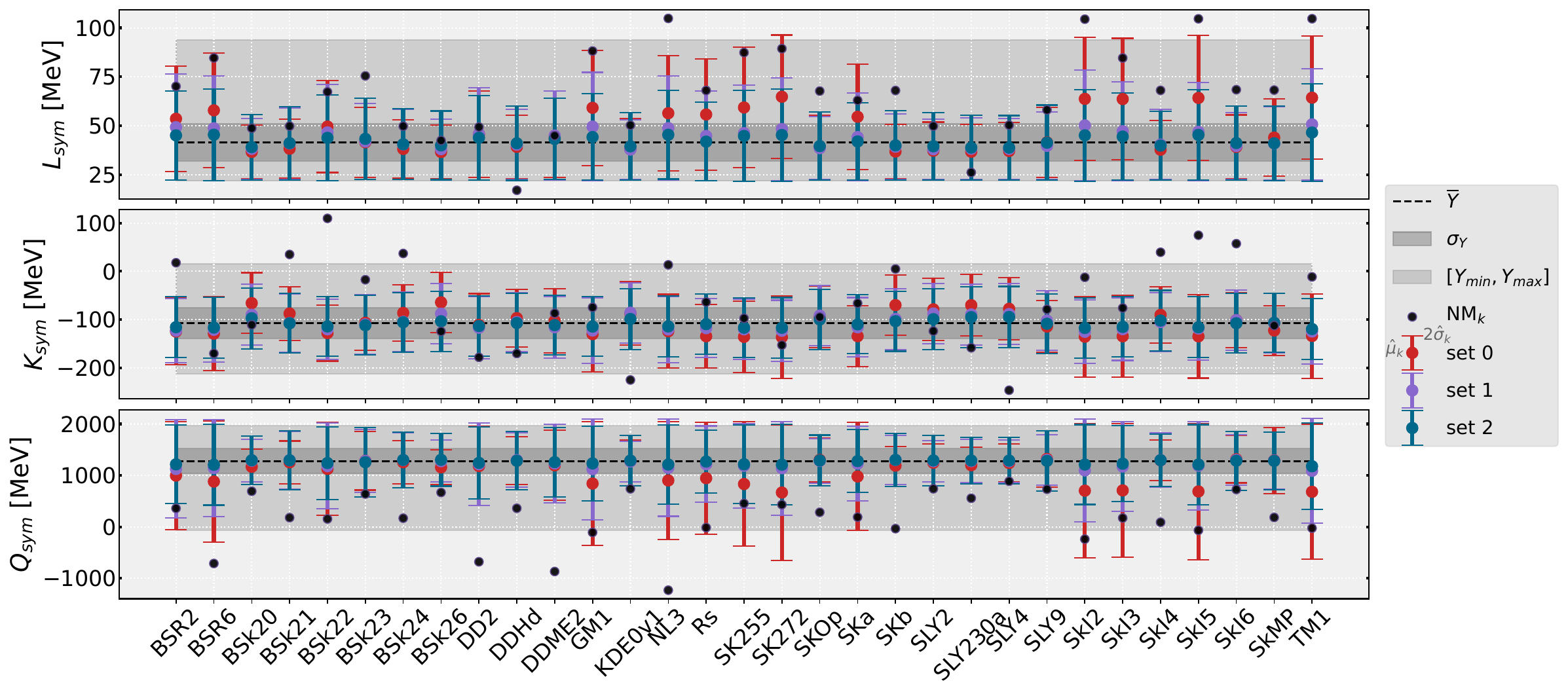}\\
    \caption{Representation of the $L_{\text{sym}},\, K_{\text{sym}},\, Q_{\text{sym}}$ output values where once again the light grey band illustrates the minimum and maximum values within the training dataset [$Y_{\text{min}},Y_{\text{max}}$], and the dashed line and dark grey band correspond, respectively, to the mean and the $1\sigma$ CI of the training region ($\overline{Y}\pm\sigma_Y$). The error bar denotes the $2 \sigma$ range prediction by the model ($\hat{\mu}_k \pm 2\hat{\sigma}_k$) for the sets 0, 1, and 2. More information can be found in the text.}
    \label{fig:NM_2}
\end{figure*}

In order to analyze the predictions obtained with the BNN models trained in the three datasets for each nuclear model,  we represent the BNN model's performance (coloured dots with error bars) and the real value of the nuclear model defined as $\text{NM}_k$ (black dots), where k represents each of the 31 nuclear models in Figs. \ref{fig:NM_1} and \ref{fig:NM_2}.

The predictions for the 100 samples of each nuclear model are presented utilizing both the law of total expectation and the law of total variance, with mean and standard deviation values denoted as $\hat{\mu}_k$ and $\hat{\sigma}_k$, respectively, where once again $k$ represents each of the 31 nuclear models. These quantities are represented by an error bar defined as $\hat{\mu}_k \pm 2\hat{\sigma}_k$ for the three datasets.
We include the mean ($\overline{Y}$) (dashed line) and standard deviation ($\sigma_Y$) of the training dataset in these figures (dark grey band). Additionally, the range of the training dataset [$Y_{min},Y_{max}$] is shown (light grey band), providing a reference for understanding the BNN model's performance relative to the training distribution.
For the higher-order properties, particularly for $Q_0$, the BNN model struggles when values significantly deviate from the training region, becoming unable to accurately track the true value. Notably, models trained with the set without uncertainty show an ability to appropriately extend the error bar, effectively capturing the true value. For certain properties, such as $Q_{\text{sym}}$, these models even reach beyond the training region.

\begin{figure}[!hbt]
    \centering
    \includegraphics[width=0.9\linewidth]{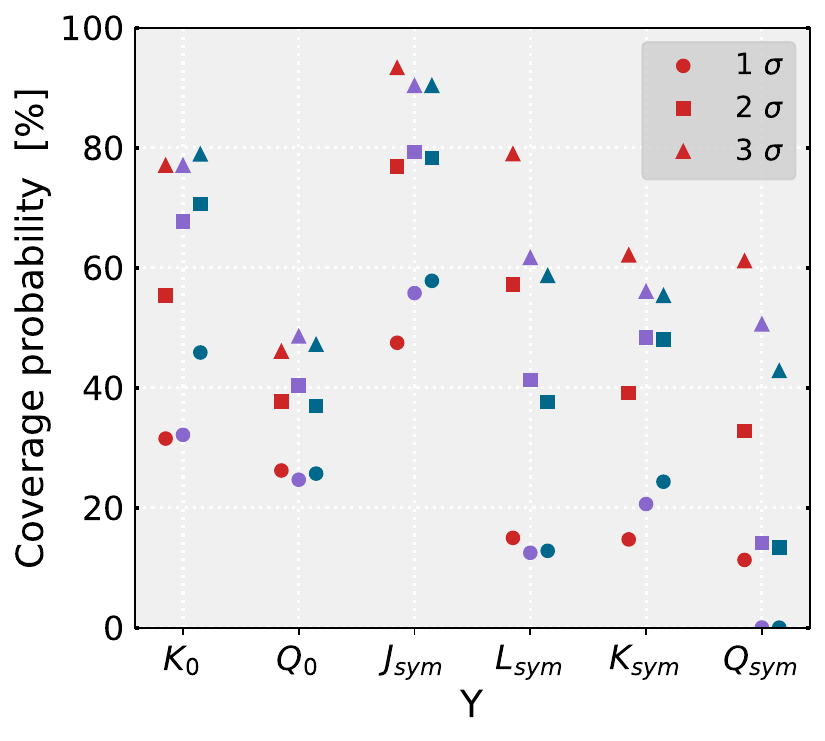}
    \caption{Coverage probability for the model tested with the three sets across 100 samples of the 31 nuclear models, with red representing set 0, purple representing set 1, and blue representing set 2.}
    \label{fig:covera_NM}
\end{figure}

To gain a clearer understanding of the BNN model's predictive performance, we have recalculated the coverage probability for 1, 2, and 3$\sigma$, as illustrated in Fig.~\ref{fig:covera_NM} for these new three datasets. This provides insight into the percentage of values our BNN model accurately captures up to $3\sigma$. Notably, $J_{\text{sym}}$ exhibits nearly 100\% coverage for $3\sigma$, indicating that the model effectively contains the majority of values within a $3\sigma$ range from the predicted mean.

When comparing the inference results between the test set of Section \ref{test_dataset} (see Fig.~\ref{fig:gamma}, right panel) and the above figure for the nuclear models (Fig.~\ref{fig:covera_NM}), it is clear a considerable decline in the BNN models performances. However, we should consider the key difference between both inferences: while the test set of Section \ref{test_dataset} has exactly the same statistical structure of the training set, these 31 nuclear models are characterized by nuclear matter properties values far from the training region; in other words, the above inferences heavily rely on the extrapolation capacity of BNN models. One obvious way of increasing the BNN models performance would be to generalize the training set to be as broad as possible, covering an extensive range of possible nuclear matter properties. However, the training set we have used was restricted to be consistent with theoretical/experimental low-energy constraints: in other words, we are not expecting the true EOS of nuclear matter, and its nuclear matter parameters, to strongly deviate from our training statistics.  Notice that the nuclear models that fail to reproduce with good accuracy the nuclear matter properties are distributed within the different frameworks used to build the EoS, i.e. the performance depends on the way the phenomenological nuclear models are constrained and not on the different frameworks, such as RMF, Skyrme forces or hadronic density dependent models. \\

\begin{figure*}[!hbt]
    \centering
    \includegraphics[width=0.9\linewidth]{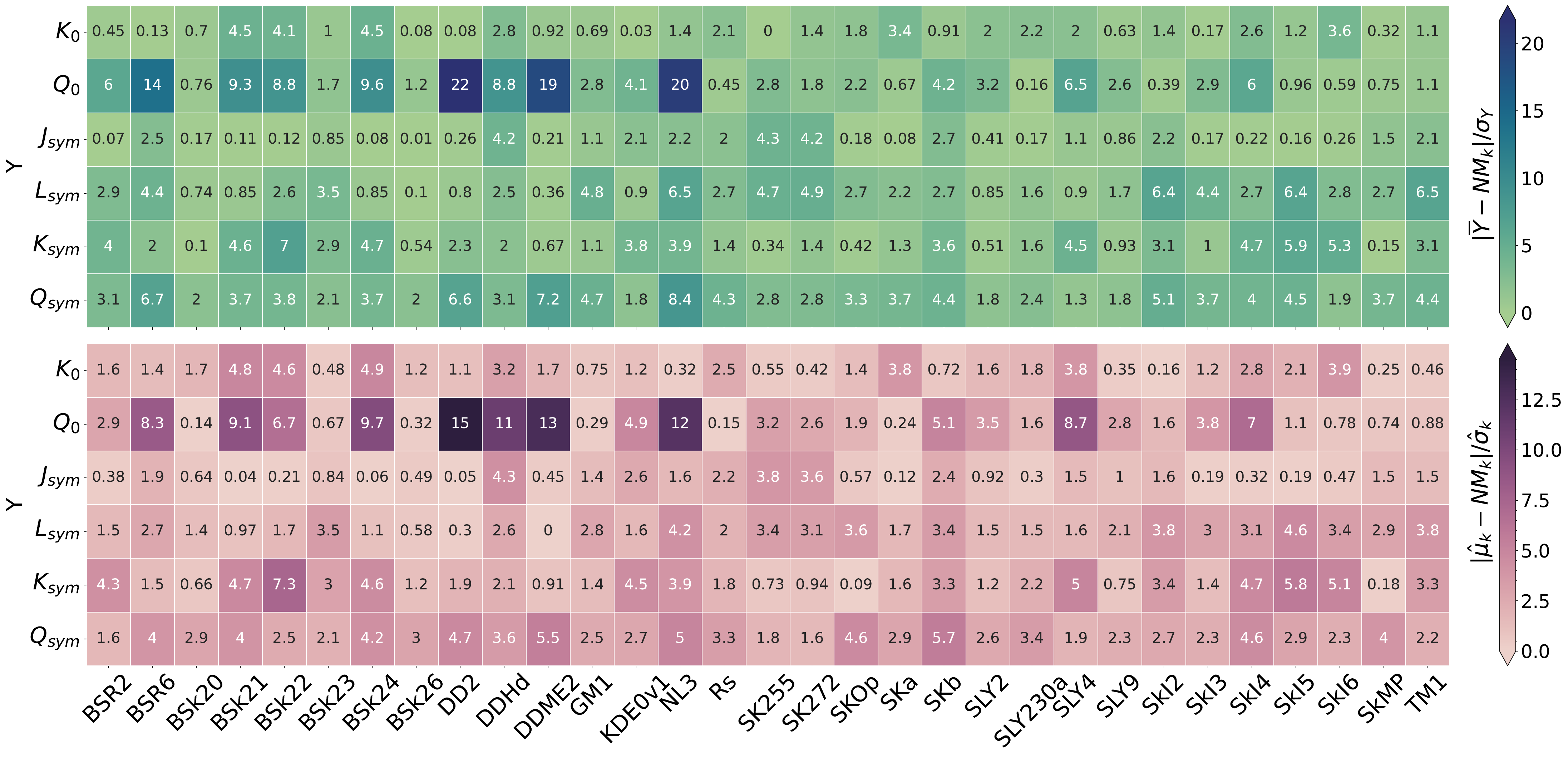}\\
    \caption{
In the top panel of the heatmap, we present an overview of the BNN model predictions' reliability, considering the normalized distance of target values (nuclear matter properties $\text{NM}_k$ for each model $k$) from the respective mean of the training region,$|\overline{Y} -\text{NM}_k|/\sigma_Y$. Here, $\overline{Y}$ and $\sigma_Y$ denote the mean and standard deviation, respectively, of the training dataset. The values $\text{NM}_k$ correspond to the nuclear matter properties, where the index $k$ represents each nuclear model.
In the bottom panel, the focus shifts to the absolute value of the normalized model residuals' predictions for set 2, and the training dataset mean and standard deviation are substituted by $\hat{\mu}_k$ and $\hat{\sigma}_k$ that symbolize, respectively, the predicted mean and standard deviation, $|\hat{\mu}_k -\text{NM}_k|/\hat{\sigma}_k$. 
For a detailed understanding of these individual quantities, see Figs. \ref{fig:NM_1} and \ref{fig:NM_2}.}
    \label{fig:heat_map1}
\end{figure*}

The performance of the BNN models is well captured by the two combined heatmaps represented in Fig. \ref{fig:heat_map1} illustrating a comparison between the training dataset properties $Y$ and the BNN model predictions $\hat{\mu}_k$ for the 31 nuclear models, with the true nuclear matter properties of these 31 EoS, respectively, in the top and bottom heatmap. 

In the upper panel, we visualize the deviation of target values, specifically the six nuclear matter properties denoted as $\text{NM}_k$ for nuclear model $k$, from the mean of the training region, $\overline{Y}$. This deviation is normalized by the standard deviation of the training region, $\sigma_Y$. Meanwhile, the lower panel portrays the absolute value of the normalized model residuals' predictions for set 2, chosen as an illustrative example. To be precise, this representation captures the distance between the true value $\text{NM}_k$ and the predicted mean for each nuclear model, $\hat{\mu}_k$, normalized by the predicted standard deviation, $\hat{\sigma}_k$. These individual quantities were shown in Figs. \ref{fig:NM_1} and \ref{fig:NM_2}. The objective of this visualization is to emphasize that nuclear model properties with a greater distance from the mean of the training data also tend to exhibit a larger deviation from the BNN predicted mean, $\hat{\mu}_k$.
This trend is evident, for instance, in the case of the quantity $Q_0$. Taking the nuclear model DD2 as an illustrative example, $Q_0$ exhibits the furthest deviation from the mean value of the training distribution and, consequently, is the farthest from the predicted mean — being 15 $\hat{\sigma}_{\text{DD2}}$ away from the predicted mean. The ranges of the mean values and standard deviations in the training set vary significantly for each quantity, as detailed in Table. \ref{tab:describe}.

In a recent study \cite{krastev2023deep}, a deep learning model was also tested with nuclear models, albeit with a different amount and output/input quantities. Notably, we utilized two common models, SK272 and SK255, specifically for the properties $L_{\text{sym}}$, $K_{\text{sym}}$, and $K_{0}$. Remarkably, we achieved smaller residuals in comparison to their results for the quantities $K_{\text{sym}}$ and $K_{0}$, despite the fact that they employed a more complex architecture and conducted a direct mapping from the EOS to the properties of nuclear matter.\\

Furthermore, in our earlier work \cite{ferreira2022extracting}, where we directly mapped the EoS of $\beta$-stable matter to the properties of nuclear matter, we also conducted tests with the same 31 nuclear models. The results were not only successful but also exhibited a higher precision. However, our main drawback in that study was the challenge of quantifying the uncertainty, a challenge that we have effectively addressed in our present work.

\subsection{Real observation dataset \label{obs_aplica}}

\begin{figure}[!hbt]
    \centering
    \includegraphics[width=1.\linewidth]{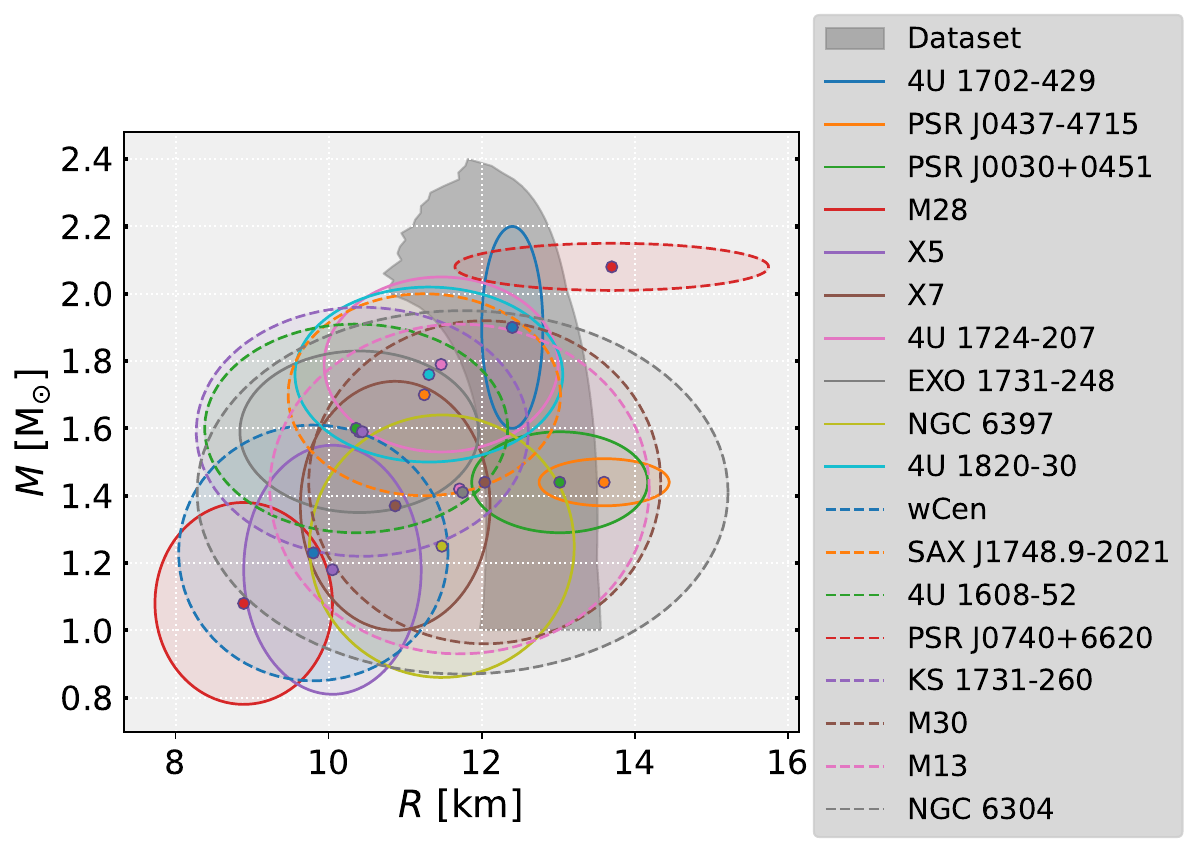}
    \caption{Observations that come from the Table \ref{tab:observat}, for the 68\% CI, the dark grey band is the dataset region without noise.  Notice that there are only three mean values inside the training region. }
    \label{fig:M_R_obser}
\end{figure}

In a final evaluation of our BNN model, we conducted a final test using the real NS observations listed in Table \ref{tab:observat}. 
Figure \ref{fig:M_R_obser} shows these observations. As their dispersion is quite high and most of them are far from the training set, we took their mean values as the observation values, which we supplied to the BNN model.\\

Contrarily to the last Section, here we employ just one BNN model to the pulsar dataset. We use a BNN model trained on set 2 because the noise properties of set 2 were determined precisely from the pulsar list, Table~\ref{tab:observat}, leading to set 2 exhibiting an input noise similar to the standard deviation of the mass-radius pairs. Additionally, this choice is motivated by the highly dispersed nature of the mean values of the pulsar observations, which fall within the domain of set 2. This implies that the model has learned these diverse regions effectively.
In a similar methodology as followed in the last section, we generate sets of 5 pairs from the 18 pulsar values through sampling without replacement, and this process was iterated 100 times, resulting in a dataset of 100 input vectors for testing. We opted for 100 repetitions to remain consistent with the previous subsection. 

Only three mean values fall within the dataset region without noise for the mass radius curves, and the dispersion of each observation, in most cases, surpasses the radius interval covered during training. This implies that the model is extrapolating values, making it challenging to obtain accurate predictions. The resulting mean and standard deviation for the six values of the output vector were obtained using, once again, the law of total variance and the law of total expectation for the 100 different input vectors. These quantities are represented as $\hat{\mu}_{obs}$ and $\hat{\sigma}_{obs}$ respectively, and are presented in Table \ref{tab:observ_prediction}. To enhance predictions, more precise observation values are crucial.

\begin{table}[!hbt]
    \caption{ The predicted mean and standard deviation, denoted as $\hat{\mu}_{obs}$ and $\hat{\sigma}_{obs}$ respectively, determined for the output vector based on the set created from the observations listed in Table \ref{tab:observat}. Further details can be found in the accompanying text.}
    \label{tab:observ_prediction}
    \centering
    \begin{tabular}{cccc}\toprule
\multirow{2}{*}{Y \ [MeV]}  & &     \multicolumn{2}{c}{Set 2}     \\ \cline{3-4} 
&& $\hat{\mu}_{obs}$& $\hat{\sigma}_{obs}$ \\
\hline
$K_0$  &  & 264.043724 &   18.184857 \\
$Q_0$ &  &-397.115823 &   55.868526 \\
$J_{\text{sym}}$ &  &  31.622006 &    1.466793 \\
$L_{\text{sym}}$ &  &  39.802926 &    8.764159 \\
$K_{\text{sym}}$& & -101.667751 &   31.432478 \\
$Q_{\text{sym}}$ & & 1294.825496 &  252.531340 \\ \hline
    \end{tabular}
\end{table}

\begin{figure*}[!hbt]
    \centering
    \includegraphics[width=0.32\linewidth]{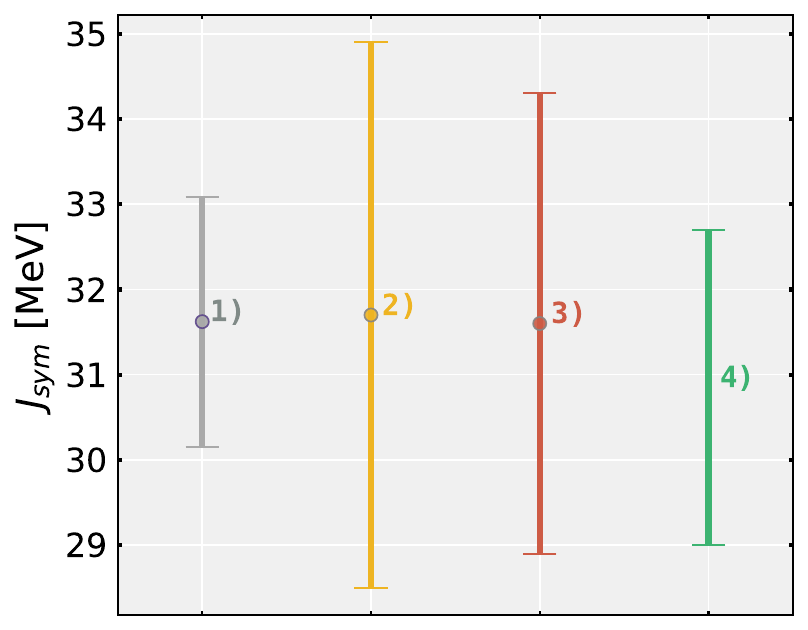}
    \includegraphics[width=0.32\linewidth]{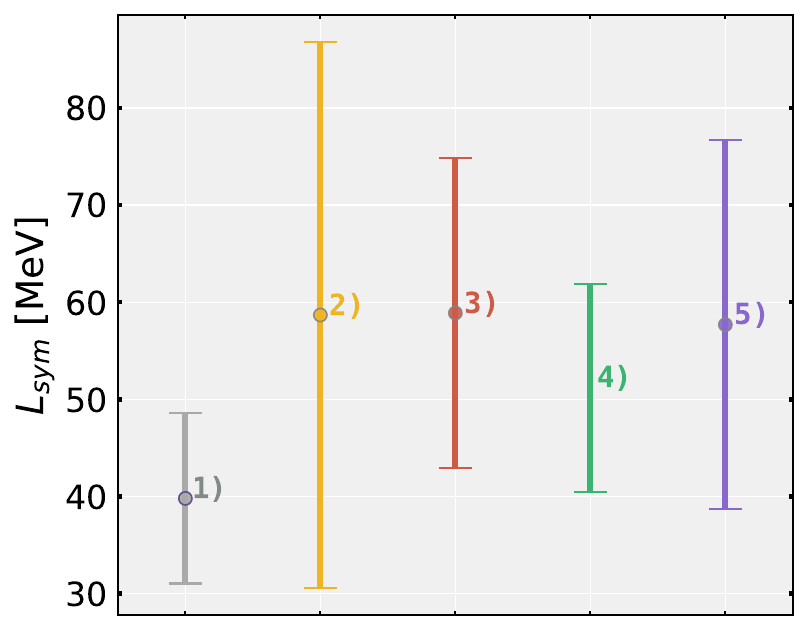}
    \includegraphics[width=0.34\linewidth]{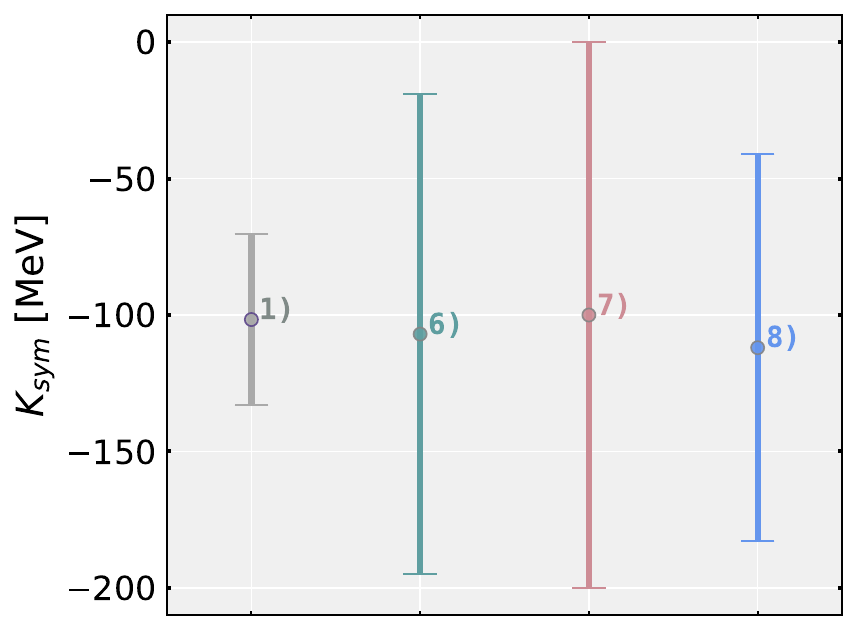}
      \caption{Comparison of our results for the symmetry energy, its slope, and curvature (labeled as 1 for a 68\% CI) with values from other studies: 2) $J_{\text{sym}}= 31.7 \pm 3.2 $ MeV and $L_{\text{sym}}= 58.7 \pm 28.1 $ MeV from \cite{Oertel:2016bki}, 3) $J_{\text{sym}}= 31.6 \pm 2.7 $ MeV and $L_{\text{sym}}= 58.9 \pm 16 $ MeV \cite{Li:2013ola}, 4) $29<J_{\text{sym}}<32.7$ MeV and $40.5<L_{\text{sym}}< 61.9$ MeV  \cite{Lattimer2013}, 5) $L_{\text{sym}}= 57.7 \pm 19 $ MeV at a 68\% confidence level \cite{Li:2021thg}, 6) $K_{\text{sym}}= -107\pm 88 $ MeV at a 68 \% confidence level in \cite{Li:2021thg}, 7) $K_{\text{sym}}= -100\pm 100 $ MeV \cite{Margueron:2017lup}, and 8) $K_{\text{sym}}= -112 \pm 71 $ MeV \cite{Mondal:2017hnh}.}
    \label{fig:compar}
\end{figure*}

The comparison between our work and existing studies for the quantities of symmetry energy, its slope, and curvature is presented in Fig. \ref{fig:compar}. Specifically, we compare our results 1) for a 68\% CI with the following references: 2) $J_{\text{sym}}= 31.7 \pm 3.2 $ MeV and $L_{\text{sym}}= 58.7 \pm 28.1 $ MeV from \cite{Oertel:2016bki}, 3) $J_{\text{sym}}= 31.6 \pm 2.7 $ MeV and $L_{\text{sym}}= 58.9 \pm 16 $ MeV \cite{Li:2013ola}, 4) $29<J_{\text{sym}}<32.7$ MeV and $40.5<L_{\text{sym}}< 61.9$ MeV  \cite{Lattimer2013} ,5) $L_{\text{sym}}= 57.7 \pm 19 $ MeV at a 68\% confidence level \cite{Li:2021thg} for 24 new analysis of neutron star data since GW170817, 6) $K_{\text{sym}}= -107\pm 88 $ MeV at a 68\% confidence level in \cite{Li:2021thg} but for the 16 new analysis of neutron star data since GW170817, 7) $K_{\text{sym}}= -100\pm 100 $ MeV \cite{Margueron:2017lup}, and 8) $K_{\text{sym}}= -112 \pm 71 $ MeV \cite{Mondal:2017hnh}.\\
The key takeaway from this comparison is that our results for the three quantities fall within the ranges reported by the seven different works, suggesting the accuracy of our model. Particularly noteworthy is the alignment of our predicted mean values with those from other studies, especially for $J_{\text{sym}}$ and $K_{\text{sym}}$. 

\section{Conclusions \label{conclusions}}

Throughout this study, we have delved into the exploration of nuclear properties, aiming to derive them directly from mass-radius observations while quantifying their respective uncertainties. Our approach involves the application of BNNs, a probabilistic machine learning model that outperforms traditional neural networks by providing uncertainty measurements alongside its predictions.
Our EOS dataset used to train the BNN model was built from a relativistic mean field approach within a Bayesian framework and incorporates crucial constraints derived from both nuclear matter properties and neutron star observations. Unlike more flexible EOS parameterizations discussed in works like \cite{Annala2019}, we deliberately opted for this specific family of microscopic nuclear models. This choice is driven by our overarching goal — to explore the feasibility of inferring nuclear matter properties, such as the quantities $K_0,\, Q_0,\, J_{\text{sym}},\,  L_{\text{sym}},\, K_{\text{sym}},\, Q_{\text{sym}}$, from observations of neutron stars. Spanning a dataset of 25 287 EOS, we generated three distinct sets of mock observational data for the mass and radius of five NS, simulating diverse scenarios of uncertainties, as outlined in Table \ref{tab:sets}. These sets represent various levels of input noise scatter. Subsequently, we trained three BNN models, each corresponding to one of the three sets.

Through this exploration, we have successfully demonstrated the efficacy of BNNs in utilizing measurements of the mass and radius from five neutron stars to extract comprehensive information about the properties of nuclear matter. Our model has demonstrated its capability to provide valuable insights while effectively addressing uncertainties, particularly for the test set crafted with the same statistical characteristics as the training set.\\

We further extended our investigation by testing the BNN model using the NS EOS derived from 31 nuclear models \cite{Fortin:2016hny}, probing its ability to predict the properties of nuclear matter when confronted with samples exhibiting a behavior vastly different from the training set. Some of these samples could fall completely outside the training region, as illustrated in Fig. \ref{fig:M-R-nuclear}. Adapting the 31 NS EOS to the three distinct sets with different uncertainties, we observed that the BNN model encountered challenges when dealing with samples that fall far away from the mean of the training region. Interestingly, we found that set 0, which included no uncertainty on the measurement of the mock data,  demonstrated a superior ability to accurately capture the target values beyond the training region, even extrapolating successfully, going towards the direction of the out of the training region values.

In a final test, we evaluated the model using a test set with real observation values, once again encountering values completely outside the training region and exhibiting a quite large dispersion. This test dataset was exclusively tested on the model trained with set 2, i.e. the set with the largest uncertainties on the mock data observations. Our key takeaway from this test is that the current available observations still require improvement, given the substantial associated uncertainty leading to limited information for the extraction of the EOS. We anticipate advancements from future observations, such as those expected from STROBE-X \cite{STROBE-X} and eXTP \cite{eXTP}, promising radius measurements with uncertainties as low as 2\%-5\%, thereby refining our predictions.\\

Numerous approaches for further exploration and improvement unfold from our current work. One potential direction involves augmenting the output vector by incorporating fourth-order components of the Taylor expansion of the nuclear matter EOS. 

One particularly promising idea for future work involves training the BNN model with a diverse range of EOS that permit the extraction of nuclear matter properties (e.g. excluding polytropics and other agnostic approaches). Specifically, this could involve training with the nuclear models tested in Section \ref{NM_aplication} and utilizing meta-models. The BNN model would be able to learn a larger domain for the input and output space. \\

Introducing hyperons or a quark phase into the composition of neutron stars and training and testing the model with this set could offer intriguing insights. 
Regarding the stochastic model, developing other prior distributions for the weights is worth for exploration.
Additionally, a crucial study would involve analyzing other distributions beyond Gaussian distributions for the posterior of the weights. Exploring alternative distributions could potentially lead to further improvements in the loss optimization. These improvements collectively offer rich possibilities for advancing and refining our current methodology.
\section*{ACKNOWLEDGMENTS} 
This work was partially supported by national funds from FCT (Fundação para a Ciência e a Tecnologia, I.P, Portugal) under the projects 2022.06460.PTDC, and 
UIDB/04564/2020 and UIDP/04564/2020, with DOI identifiers 10.54499/UIDB/04564/2020 and 10.54499/UIDP/04564/2020, respectively.

\onecolumngrid
\appendix

\section{Correlations\label{ap:correl}}

\begin{figure}[!hbt]
    \centering    \includegraphics[width=0.4\linewidth]{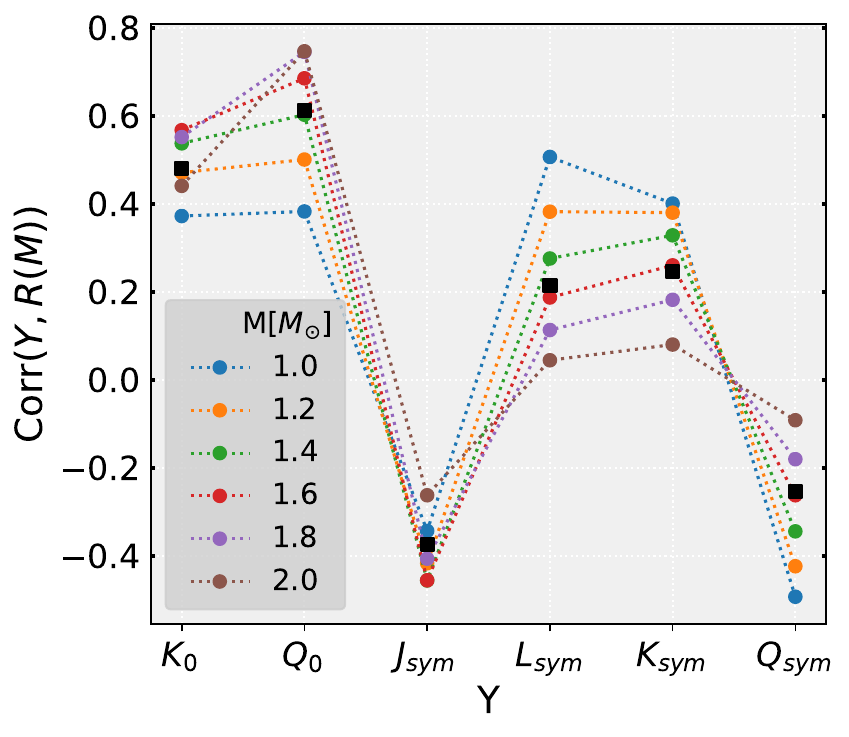}
    \caption{Correlation between the nuclear matter parameters and the radius for fixed mass values $R(M)$. In black squares we show the mean correlation value for $M/M_{\odot}\in [1,2]$.}
    \label{fig:corre}
\end{figure}

Let us calculate and analyze the Pearson correlation, for the source dataset, between the parameters of nuclear matter and the corresponding radius values for fixed mass values. While it is a limited quantity as it only measures linear correlations between variables, it already gives a good indication of the important properties in describing the NS radius. The results are shown in Fig. \ref{fig:corre}. The most notable observation is that the quantity $Q_0$ exhibits the highest correlation coefficient. By considering the isovector nuclear matter properties, $\{J_{\text{sym}},...,Q_{\text{sym}}\}$, it is clear an interesting feature: the low order parameters are more strongly correlated with the low $M$ while the opposite happens for high order parameters. This is expected since low NS masses have central densities much smaller than massive NS and thus low order parameters are enough to explain the radii of light NS. On the other hand, massive NS radii are more sensitive to high order parameters because their central densities are very sensitive to the high order polynomial orders in the EOS. This plot serves as a valuable tool for comprehending the results of this work.  A deeper study on the subject of correlation can be found in \cite{Ferreira:2019bgy}.

\section{Table of observations}

Illustration of the observation table outlined in the article \cite{Soma:2022vbb}, featuring 18 observables derived through Gaussian-fitted analyses. These observations are established from marginalized distributions, capturing key properties associated with neutron stars.

\begin{table}[!hbt]
\caption{
Observation table sourced from the article \cite{Soma:2022vbb}.}
\label{tab:observat}
\begin{tabular}{lccccc}
\toprule
\multirow{2}{*}{Observables} &\multicolumn{2}{c}{M ($\textup{M}_\odot$)} & &\multicolumn{2}{c}{R (Km)}       \\\cline{2-3} \cline{5-6} 
      &     Mean & $\sigma$ & &      Mean& $\sigma$ \\ \hline
     4U 1702-429 &   1.9 &    0.3 &  &  12.4 &    0.4 \\
  PSR J0437-4715 &  1.44 &   0.07 &   &13.6 &   0.85 \\
  PSR J0030+0451 &  1.44 &   0.15 &  &13.02 &   1.15 \\
             M28 &  1.08 &    0.3 &  & 8.89 &   1.16 \\
              X5 &  1.18 &   0.37 &  &10.05 &   1.16 \\
              X7 &  1.37 &   0.37 &  &10.87 &   1.24 \\
     4U 1724-207 &  1.79 &   0.26 &  &11.47 &   1.53 \\
    EXO 1731-248 &  1.59 &   0.24 &  & 10.4 &   1.56 \\
        NGC 6397 &  1.25 &   0.39 &  &11.48 &   1.73 \\
      4U 1820-30 &  1.76 &   0.26 &  &11.31 &   1.75 \\ \hline
            wCen &  1.23 &   0.38 &  &  9.8 &   1.76 \\
SAX J1748.9-2021 &   1.7 &    0.3 &  &11.25 &   1.78 \\
      4U 1608-52 &   1.6 &   0.31 &  &10.36 &   1.98 \\
  PSR J0740+6620 &  2.08 &   0.07 &  & 13.7 &   2.05 \\
     KS 1731-260 &  1.59 &   0.37 &  &10.44 &   2.17 \\
             M30 &  1.44 &   0.48 &  &12.04 &    2.3 \\
             M13 &  1.42 &   0.49 &  &11.71 &   2.48 \\
        NGC 6304 &  1.41 &   0.54 &  &11.75 &   3.47 \\
\hline
\end{tabular}
\end{table}

\twocolumngrid
\bibliographystyle{apsrev4-1}
%\bibliography{biblio}
%

\end{document}